\documentclass[sigconf]{acmart}
\AtBeginDocument{%
  }

\setcopyright{acmlicensed}
\copyrightyear{2018}
\acmYear{2018}
\acmDOI{XXXXXXX.XXXXXXX}
\acmConference[Conference acronym 'XX]{Make sure to enter the correct
  conference title from your rights confirmation email}{June 03--05,
  2018}{Woodstock, NY}
\acmISBN{978-1-4503-XXXX-X/2018/06}

\usepackage{tabularx}
\usepackage{multirow}
\usepackage{makecell}
\usepackage{array}

\begin{document}

\title{An Interactive LLM-Based Simulator for Dementia-Related Activities of Daily Living}

\author{Kruthika Gangaraju}
\email{kgangaraju@wpi.edu}
\affiliation{%
  \institution{Worcester Polytechnic Institute}
  \city{Worcester}
  \state{Massachusetts}
  \country{USA}
}

\author{Shu-Fen Wung}
\email{swung@health.ucdavis.edu}
\affiliation{%
  \institution{University of California Davis}
  \city{Sacramento}
  \state{CA}
  \country{USA}}

\author{Kevin Berner}
\email{kberner@mghihp.edu}
\affiliation{%
  \institution{MGH Institute of Health Professionals}
  \city{Boston}
  \state{MA}
  \country{USA}}

\author{Jing Wang}
\email{jing.wang1@unh.edu}
\affiliation{%
  \institution{University of New Hampshire}
  \city{Boston}
  \state{MA}
  \country{USA}}
  
\author{Fengpei Yuan}
\email{fyuan3@wpi.edu}
\affiliation{%
  \institution{Worcester Polytechnic Institute}
  \city{Worcester}
  \state{Massachusetts}
  \country{USA}
}

\renewcommand{\shortauthors}{Gangaraju et al.}

\begin{abstract}
Effective dementia caregiving requires training and adaptive communication strategies, yet assistive technologies (e.g., AI and robotics) for dementia care are limited by the scarcity of context-rich, privacy-sensitive data capturing how people living with Alzheimer’s disease and related dementias (ADRD) behave during activities of daily living (ADLs). We present a web-based simulator that leverages a large language model (\texttt{gpt-5-mini}) to generate multi-turn, stage and care-setting-conditioned behaviors for a simulated person living with ADRD during ADL assistance, combining verbal utterances with lightweight nonverbal/behavioral cues (in parentheses). Users configure dementia severity, care setting (and time in setting), and ADL; after each simulated patient turn, users provide a 1--5 realism rating with optional critique, then respond as the caregiver via free text or by selecting/editing one of four strategy-scaffolded suggestions (\textit{Recognition}, \textit{Negotiation}, \textit{Facilitation}, \textit{Validation}). We conducted a formative domain-expert-in-the-loop evaluation deployed online (14 dementia-care experts; 18 sessions; 112 rated turns). Overall, simulated PLWD behavior was judged moderately to highly plausible, with 
scenario-dependent variability and a typical session length of six turns. Expert participants most often authored custom responses (54.5\% of turns); among strategy-scaffolded suggestions, \textit{Recognition} and \textit{Facilitation} were used frequently. Thematic analysis of domain experts' critiques yielded a six-category failure-mode taxonomy, highlighting recurrent breakdowns in ADL task grounding and care-setting consistency and informing concrete prompt and workflow refinements. The simulator and interaction history support an evidence-driven refinement loop toward validated patient-caregiver co-simulation and provide a foundation for data collection, caregiver training, and assistive AI/robot policy development.
\end{abstract}



\received{20 February 2007}
\received[revised]{12 March 2009}
\received[accepted]{5 June 2009}

\maketitle

\section{Introduction}
Alzheimer’s disease and related dementias (ADRD) progressively impair cognitive functions such as memory, attention, language, and executive function \cite{AA2024diseasefact} \cite{langbaum2023recommendations}. These impairments  often manifest most apparently during activities of daily living (ADLs) such as toileting, dressing, bathing/showering, transferring, taking medicine, and eating \cite{giebel2014deterioration}. These moments are safety-critical and impacted by challenging behaviors \cite{yin2023risk, spira2006behavioral}. Persons living with ADRD (PLWDs) may become confused about the goal of the task, lose the sequence of steps, misunderstand caregiver prompts, or respond with reactions of agitation, withdrawal, repetition, or confabulation. In the field of technology-assisted dementia care—spanning assistive robotics, context-aware prompting, and AI-based decision support—modeling these behaviors is essential \cite{ghafurian2021social, holderautomated}. Yet, there exists a notable scarcity of robust, context-rich behavioral data describing \textit{what a person might communicate and how one might behave} at different stages of dementia during ADLs \cite{pappada2021assistive}. The collection of such data in real-world settings is costly and time-consuming, further constrained by privacy concerns, the need for informed consent, and the practical difficulties of capturing rare but important events (e.g., unsafe actions, escalation of behaviors, or refusal). This lack of scalable, staged, and context-specific behavioral data is a major bottleneck for developing, testing, and benchmarking ADL assistance technologies.

Large language models (LLMs) offer a new opportunity for generating plausible, context-conditioned behavioral simulations from textual descriptions \cite{park2023generative, argyle2023out}. In principle, an LLM can produce a diverse, stage-appropriate patient utterances and behaviors within ADL contexts, enabling rapid scenario creation for prototyping, training, and evaluation. However, a central challenge is the validity of these simulations \cite{hu2025simbench}.  Without careful grounding and human verification, simulated behavior may be unrealistic, stereotyped, or clinically implausible-leading to models, robots, or AI agents that "perform well" in simulation environments but fail in real care settings \cite{qu2024performance, huang2025survey}. To be useful to researchers, clinicians, and educators, an LLM-based dementia ADL simulator must demonstrate two critical criteria: (1) that the generated behaviors align with expectations for the selected dementia stage and context, and (2) that they support realistic caregiver decision-making about appropriate responses to assist the person in completing the task safely and respectfully.

\begin{figure}[ht!]
  \centering  \includegraphics[width=\columnwidth]{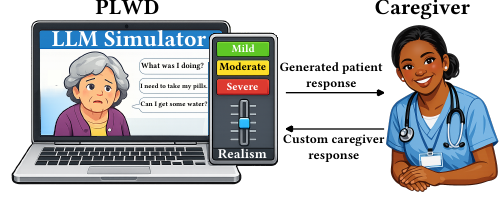}
\caption{Simulated person with Alzheimer’s dementia (PLWD) interacting with a human caregiver at a selected dementia stage.}
 \label{fig:simulatorIllustration}
\end{figure}
In this work, we present an online, interactive Alzheimer’s ADL simulator (Fig.~\ref{fig:simulatorIllustration}) that uses an LLM to generate responses multi-turn ADL interactions for a simulated person living with dementia during ADL assistance. Users configure the dementia stage (e.g., early, moderate, or late), care setting, and ADL scenario; the simulator then generates verbal utterance paired with nonverbal/behavioral cues (in parentheses) to support stage- and context-conditioned interaction.

We conducted a formative, domain-expert-in-the-loop evaluation to assess whether the generated behavior are realistic. At each turn, participants provide a realism rating (1--5) with optional critique and respond as the caregiver either by authoring free text or by selecting/editing one of four strategy-scaffolded suggestions (Recognition, Negotiation, Facilitation, Validation). Our evaluation is guided by three research questions: (RQ1) To what extent do LLM-generated behaviors align with the selected dementia stage and care setting? (RQ2) How useful are the strategy-based caregiver suggestions for supporting realistic interactions? (RQ3) What recurrent failure modes emerge, and what design requirements do they imply for improving future iterations?



Beyond providing an evaluation workflow, this simulator enables scalable collection of two complementary signals: (i) human judgments about realism and breakdowns in simulated patient behavior, and (ii) in-context caregiver responses conditioned on specific patient states within ADL scenarios. Together, these data support an evidence-driven refinement loop for prompt/workflow updates and create a foundation for downstream work such as scenario-based training, benchmarking across ADLs and stages, and learning safer assistance policies for interactive AI/robotic systems. More broadly, we view a validated, configurable simulator as a practical step toward digital-twin-style modeling of dementia-related ADL interactions, enabling systematic "what happens next" exploration under different caregiver actions.

This paper presents the following contributions:
\begin{enumerate}
    \item An interactive, stage-conditioned ADL simulator that leverages an LLM to generate multi-turn patient behavior (verbal and described nonverbal/behavioral cues) specifically tailored to various selected dementia stages, living situations (care setting), and ADL contexts.
    \item A scalable evaluation protocol developed to collect domain-expert judgments regarding the realism of simulated patient behavior. This protocol includes optional qualitative critiques to identify potential failure modes (e.g., stage mismatch, implausible actions, unrealistic language or scenarios).
    \item An in-context caregiver response dataset capturing the strategies used by caregivers during each interaction step, via both structured option selections and open-ended, user-authored responses.
    \item An analysis framework designed to improve simulation realism and usefulness. This framework supports future work on digital-twin-style modeling, assistive robotics/AI policy development, and dementia care workforce education and training.
\end{enumerate}

\subsection{Related Work}

\textbf{Healthcare dialogue simulators.} A growing body of work uses conversational simulations to scale communication practice in healthcare and caregiving, motivated by the high costs, variability, and limited availability of human role-players. Early systems have demonstrated that patient-like dialogue agents can support deliberate practice accompanied by structured feedback \cite{brugge2024large, sehgal2025pal}. For example, ClientBot is a text-based "patient-like conversational agent" designed to train basic counseling skills \cite{tanana2019development},
illustrating how interactive simulated dialogue can complement traditional instruction by enabling repeated practice and measurable skill outcomes.

\textbf{Dementia-specific communication training.} In dementia care, how caregivers communicate can be as critical as what they do, particularly during breakdowns nad emotionally-charged moments.Conversation Analysis--Based Simulation (CABS) proposes a training approach grounded in conversation analysis and empirical observations of real interactions to build simulations that target practical communication improvements (e.g., reducing escalation, supporting personhood, and managing breakdowns) \cite{pilnick2023conversation}. This line of work motivates dementia simulation design that captures interactional functions (not simply providing “correct answers”) and treats realism in terms of social/interactional fidelity. However, workshop-style approaches can be less scalable without computational generation and configurable scenario control.

\textbf{LLM-enabled virtual patients and feedback.} Recent studies show that LLMs can substantially increase the flexibility and scope of virtual patients by supporting open-ended questions, contextualized responses, and automated coaching~\cite{cook2025virtual,luo2025large}. \cite{cook2025virtual} evaluates an LLM-driven virtual patient for clinician--patient dialogue practice with feedback, comparing it against fixed-response chatbots and human standardized patient encounters.
\cite{luo2025large} introduces an LLM-based digital patient system built from anonymized real clinical data, reporting improved learner performance and correlations between automated scoring and human assessment, while also highlighting limitations relevant to realism (absence of a physical body) and privacy-preserving deployment constraints.
Beyond text-only interactions, systems integrating LLMs with embodied conversational agents in immersive environments (e.g., VR) emphasize that realism in communication training includes handling topic drift and unpredictable turns \cite{zhu2025designing}, which are common in dementia-oriented interactions.






\textbf{Summary and gap.} Across these efforts, simulations have progressed from scripted or domain-specific dialogue agents to LLM-enabled virtual patients with feedback and, increasingly, embodied/immersive interaction. However, most LLM virtual-patient work focuses on clinical interviewing or history-taking rather than \emph{caregiving interactions during ADLs}, and dementia-oriented training approaches emphasize interactional principles but are less often operationalized as configurable, stage-conditioned, scalable simulators that can be evaluated and iteratively refined. This motivates a dementia ADL simulator that 
(1) generates multi-turn behaviors characteristic of dementia (including verbal and nonverbal cues) conditioned on dementia severity and care setting, (2) collects turn-level human judgments of realism and failure modes, and
(3) captures in-context caregiver responses via both strategy scaffolds and free text.
Such a simulator platform supports iterative improvement and facilitates future patient--caregiver co-simulation, downstream assistive AI/robotics policy development, and dementia care workforce education.


\begin{figure*}[t]
  \centering  \includegraphics[width=0.75\textwidth]{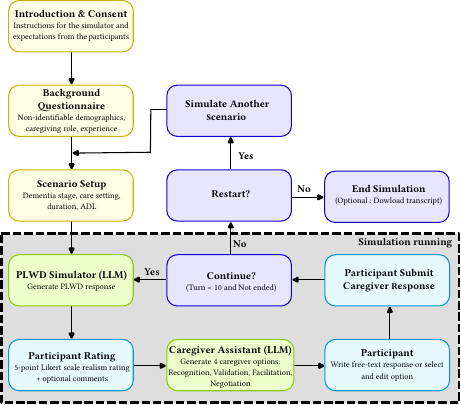}
\caption{
End-to-end participant-facing workflow of the web-based dementia–ADL simulator. After consent and a brief background questionnaire, participants configure a scenario (dementia stage, care setting, ADL). Within the dashed region (turn-level interaction), the LLM-based patient simulator generates a PLWD response; participants provide a realism rating and optional critique, then respond as the caregiver either via free-text or by selecting a strategy-scaffolded suggestion (Recognition, Validation, Facilitation, Negotiation; edits optional). The loop continues until the participant terminates the scenario or reaches turn 10. Participants may then reset the scenario to start a new simulation or end the session. Interaction transcripts can be downloaded.}
 \label{fig:simulator}
\end{figure*}

\section{Simulator System Design}
The simulator was designed around six key requirements: (R1) controllable scenarios, including the dementia stage, care setting, and ADL task; (R2) task grounding via stepwise ADL progression to maintain context and realism; (R3) lightweight, in-the-loop expert rating and feedback collection; (R4) provision of plausible caregiver response suggestions through structured strategy scaffolds; (R5) full instrumentation and exportation of interaction logs for analysis; and (R6) secure separation of API access keys to protect the model's integrity and participant data. These requirement informed the workflow and architecture described below.

\subsection{User Flow and Simulation Interface}
We designed a web-based simulator to organize expert/caregiver-PLWD interactions into a structured, repeatable, instrumented workflow, as demonstrated in Figure \ref{fig:simulator}. Participants (individuals with dementia care experience) first view an on-screen introduction describing the simulator’s purpose and providing operating instructions, and then provide consent to participate. Next, participants complete a brief background questionnaire capturing information about professional role and experience in dementia caregiving. 

Participants then configure a scenario by selecting dementia severity (stage), care setting (including the duration spent in the care setting), and an ADL. Optional free-text fields allow them to specify "Other" selections and task-specific challenges. Once the scenario is launched, the simulator runs a turn-based loop: the system generates the response from the PLWD, conditioned on the chosen scenario and task context.  Participants immediately rate the realism of the generated PLWD response on a 1–5 scale and may provide an optional qualitative comment/critiques.  After this, they respond as the caregiver by either writing free text or selecting from four LLM-generated caregiver suggestions grounded in distinct dementia communication strategies (Recognition, Negotiation, Facilitation, Validation). The selected suggestion can be edited before submission. Throughout the interaction, the conversation history remains visible to support continuity.  

The interaction continues until the participant ends the session, restarts with a new scenario, or reaches a predefined maximum number of turns (which can be configured in the simulator). All interaction data—including scenario parameters, generated patient responses, realism ratings and critiques, selected/edited caregiver responses, and timestamps—are logged for analysis. At the end of the session, participants can export transcript of the interaction. To ensure security, model access is handled via a server-side proxy to avoid exposing API credentials in the browser.

\subsection{Person with Dementia Simulation}
We simulate the responses of individuals living with ADRD using an LLM from the GPT-5 family. In our deployment, we use the \texttt{gpt-5-mini}; the model endpoint can be readily swapped if necessary. The simulator is designed to generate behavior that is consistent with dementia stage, care setting, and ADL context, including both verbal utterance and subtle nonverbal cues (expressed through parenthetical actions or emotional expressions). 

\begin{figure*}[t]
  \centering
  \includegraphics[width=0.7\textwidth]{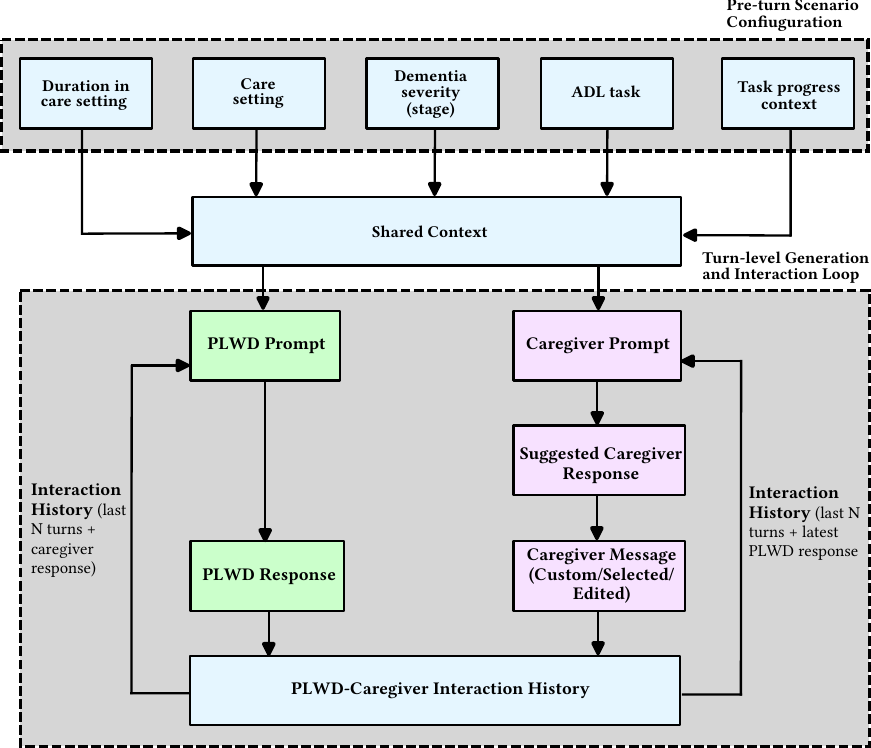}
\caption{Prompt design and turn-level context flow for generating PLWD behaviors and caregiver responses. Scenario configuration (care setting and time in setting, living situation, dementia severity, ADL, and task progress context) forms a shared context. Each turn uses a windowed interaction history to condition (i) PLWD response generation and (ii) strategy-scaffolded caregiver response suggestions.}
 \label{fig:architecture}
\end{figure*}

Each simulated patient turn is generated from a structured prompt that includes: (i) a role specification (an older adult living with ADRD  at the selected severity), (ii) scenario context (dementia severity level, care setting, and duration in that setting), (iii) the selected ADL, (iv) a task-progress scaffold (current and next ADL substep), (v) a \emph{windowed} interaction history (recent turns only) and the most recent caregiver message, and (vi) formatting constraints (e.g., 1-3 sentences; no explicit labels), as shown in Figure \ref{fig:architecture}. The windowed history improves controllability and limits prompt growth while preserving short-horizon coherence. To support reproducibility and basic security, patient generation is invoked via a backend proxy endpoint (preventing exposure of API credentials in the browser). See Section~\ref{Sec:SystemImplementation} for implementation details.

\subsubsection{Alzheimer's Dementia Severity (Stage)}
Participants select one of three dementia severity settings: Early, Middle or Late stage. In this version of the simulator, severity is operationalized using Alzheimer’s disease staging definitions. Although ADRD is broader, we restrict symptom modeling to Alzheimer’s disease because behavioral and communication profiles vary substantially across dementia etiologies (e.g., Parkinson’s disease dementia, DLB, FTD)~\cite{alz_facts_figures_2025}, and we aim to control heterogeneity in this formative evaluation. Each stage is defined by specific, prompt-level behaviors affecting memory, language, orientation, and functional dependence, based on clinical evidences \cite{alz_facts_figures_2025, mayo_alzheimers_stages_2025}:

\paragraph{{Early-stage (mild) Alzheimer's}}
Individuals exhibit predominantly short-term memory lapses and word-finding difficulty. They may occasionally misplace items or repeat themselves and experience mild strain in executive function (planning/organizing). Typically, they can maintain independence in basic ADLs but benefit from gentle reminders, simplified structure for complex tasks, and reassurance when corrected.
\paragraph{{Middle-stage (moderate) Alzheimer's}} There is increased disorientation to time/place, greater memory gaps (including forgetting personal details), and disrupted language/thought processes (confused words, confabulation-like fill-ins). Individuals often require step-by-step prompting and assistance with instrumental ADLs and some basic self-care tasks.  They may show agitation, suspicion, wandering, sundowning patterns, or intermittent refusal of care.
\paragraph{{Late-stage (severe) Alzheimer's}} Individuals show marked loss of situational awareness, minimal initiation of activities, and limited meaningful conversation. Communication often becomes sparse (single words/sounds) and may be nonverbal.  They are fully dependent on others for basic ADLs and may exhibit discomfort via their affect/behavior. Responses are primarily reactive, and interactions benefit from very simple one-step guidance, a calm tone, and familiar cues.

\subsubsection{Care setting and its duration.}
The care setting provides an environmental context that modulates PLWD's familiarity, disorientation, and behavioral affect (e.g., navigating an unfamiliar facility compared to following a long-standing home routine). We additionally capture the time spent in the selected setting to estimate acclimatization (e.g., recent transition versus long-term residence) and adjust condition accordingly.

\subsubsection{Activity of daily living and task grounding.}
Participants choose one ADL from a predefined set (with an \emph{Other} option for free-text tasks). For each predefined ADL, we encode a stepwise task plan and compute a per-turn task-progress context (current and next step) that is appended to the prompt. This scaffold anchors the model to goal-directed progression and reduces off-topic drift while still allowing for realistic breakdowns (e.g., forgetting, refusal, confusion), that are consistent with the selected stage. The patient prompt enforces concise responses (1--3 sentences) and requests nonverbal cues only when appropriate, indicated in parentheses. The user interface (UI) distinctly displays this parenthetical content, enabling lightweight behavioral annotation without additional labeling.

\subsection{Caregiver Response Simulation}
To scaffold realistic caregiver utterances and enable comparable expert interactions dementia care expert interactions across various scenarios, the simulator (\texttt{gpt-5-mini}) generates candidate caregiver responses grounded in four commonly used person-centered dementia-care communication strategies: \textit{Recognition, Negotiation, Facilitation,} and \textit{Validation}~\cite{savundranayagam2015language, o2022expanding}. At each turn, the system considers the selected dementia stage, care setting, ADL, task-progress context, and the most recent PLWD utterance to produce one concise caregiver option per strategy (typically 1--2 sentences), as illustrated in Figure \ref{fig:architecture}. 

These options are presented alongside an editable free-text input. Participants can either write their own response or click a suggested option to populate the textbox and edit it before sending. This design serves two purposes: (i) it reduces participant burden by providing strategy-aligned starting points, and (ii) it enables systematic analysis of which strategy templates experts accept, modify, or reject across different stages and tasks.
Because these strategy labels are central to both the interaction design and our subsequent analysis, we define below how we operationalize each strategy, grounded in commonly used person-centered dementia-care practices~\cite{savundranayagam2015language, o2022expanding}.

\subsubsection{Dementia Caregiving Strategies}
We operationalize the four strategies as follows, focusing on their primary functions in-context:
\paragraph{{Recognition} (\emph{personhood/identity})}
This involves interactional moves that treat the PLWD as a unique individual rather than a set of care tasks. Indicators of this approach include using the person’s preferred name/title, giving individualized greetings, affirming or rephrasing to show attentive engagement; and (when appropriate) referencing, enduring, preferences, roles, or biographical details to provide context for the current moment.
\paragraph{{Negotiation} (\emph{choice/agency})}
This strategy includes moves that consult the PLWD’s preferences and incorporate them into the care plan. Indicators include offering bounded choices, asking for permission, checking readiness, adjusting plans in response to refusals, and maintaining continuity by referencing previously expressed preferences when available.
\paragraph{{Facilitation} (\emph{supported participation})}
This involves scaffolding that enables task completion while preserving the PLWD’s participation. Indicators of this strategy include breaking ADLs into manageable steps, providing prompts/ modeling, pacing the activities, arranging the environment (e.g., placing items within reach), and offering supportive affirmations. This also includes making \emph{intention to fulfill} statements to acknowledge a request and assure the individual that it will be addressed after completing the current task, providing predictability and reassurance.
\paragraph{{Validation} (\emph{emotion-level attunement})}
Empathy-focused responses that acknowledge and legitimize the PLWD’s affective state, particularly when content is ambiguous, repetitive, or misaligned with reality. Indicators include naming/mirroring emotion, conveying understanding, and prioritizing relational comfort over factual correction.

In our strategies, \textit{Recognition} foregrounds identity, \textit{Negotiation} foregrounds autonomy, \textit{Facilitation} foregrounds guided task engagement, and \textit{Validation} foregrounds emotional alignment. Although multiple functions may co-occur simultaneously within a single utterance, we treat each generated option as exemplifying a primary strategy to support consistent presentation and facilitate subsequent analysis.

\subsection{Logging and Instrumentation}
\paragraph{PLWD-side logs.}
For each turn, we log the scenario configuration (dementia severity level, care setting and duration in the care setting, ADL), the task-progress context, the windowed PLWD-caregiver interaction history used for generation, the generated PLWD response (including verbal utterance and non-verbal cues in parentheses), and timestamps. In addition, we record expert realism ratings for each turn and optional qualitative comments immediately following each patient response.  This allows us to conduct both quantitative and qualitative analyses of stage congruence and failure modes across different conditions.

\paragraph{Caregiver-side logs.}
For each turn, we log the following information: (i) all four generated strategy-scaffolded caregiver options, (ii) the option (if applicable) selected by the participant, (iii) any edits made to the selected text  (captured as the final submitted caregiver message), and (iv) timestamps. This enables downstream analyses of strategy preferences and editing patterns across different dementia stages, care settings, and ADLs.  It also supports the iterative refinement of the caregiver-response generation module.

\begin{figure*}[t]
  \centering
  \includegraphics[width=\textwidth]{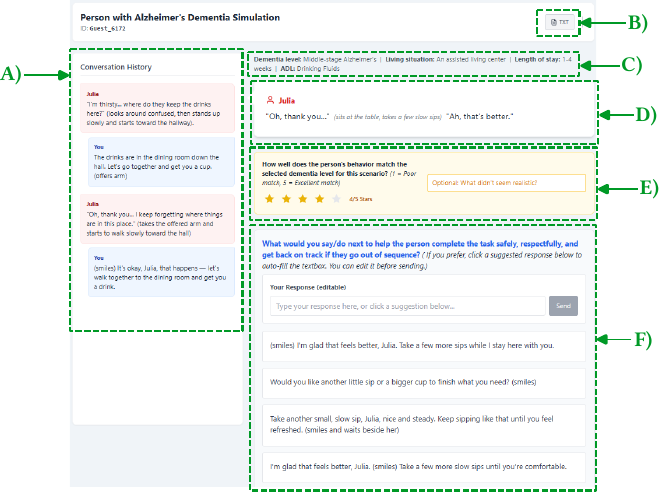}
\caption{Simulation page as displayed to participants during interaction with the simulated PLWD. A) Conversation history updates as the interaction proceeds; B) Downloadable transcript of the entire interaction for participants records; C) Simulation settings chosen by the participant; D) Generated PLWD response (black text: verbal utterance; grey text in parentheses: nonverbal behavior); E) Realism rating (1-5) for the current generated PLWD response, with optional free-text critique; F) Caregiver response interface: participants may write a free-text response or select a strategy-scaffolded suggestion.}
 \label{fig:overview}
\end{figure*}

\section{System Architecture and Implementation}~\label{Sec:SystemImplementation}
The end-to-end architecture of the simulator is implemented as a lightweight client--server web application with cloud-based logging. The system comprises (1) a browser-based client for study workflow and turn-level capture, (2) a Flask backend proxy for secure LLM access, and (3) a Firestore logging layer for persistent session/turn storage. At each turn, the client submits scenario state and dialogue context to the proxy, receives generated PLWD text and strategy options, then logs outputs and expert judgments to Firestore. The design separates (1) the interactive study interface used by participants, (2) secure model access for PLWD response generation and strategy-scaffolded caregiver suggestions, and (3) persistent storage for analysis and export. This separation supports reproducibility of the evaluation pipeline while minimizing exposure of sensitive credentials and enabling structured data capture at the granularity of individual turns.

\subsection{Client Application}
The participant-facing interface is implemented as a single-page web application (SPA) running in the browser. The SPA implements a stepwise study workflow -- (1) introduction and consent, (2) caregiving background survey, (3)scenario configuration, and (4) a turn-based simulation loop (Fig.~\ref{fig:overview}).  The application displays the full conversation history for continuity and collects turn-level expert judgments (realism ratings and optional comments provided by participants) immediately after each generated PLWD response. The application also mediates caregiver interaction by presenting strategy-conditioned suggestions alongside an editable or custom free-text input.  This allows participants to select, modify, and submit caregiver utterances. Throughout the session, the application maintains the local interaction state (e.g., current turn index, scenario configuration, and transcript) and exposes controls to advance turns, restart with a new scenario, end the interaction, and export the final transcript for record-keeping.

Within the simulation-loop view (Fig.~\ref{fig:overview}), the interface was intentionally designed to minimize navigation and cognitive load by co-locating (i) the running context (conversation history), (ii) scenario settings, (iii) the generated PLWD turn, and (iv) turn-level evaluation and response tools, enabling dementia care experts (e.g., clinicians) to rapidly rate realism and respond within the same view.

\subsection{Backend Proxy and Model Access}
All LLM calls are routed through a server-side proxy implemented in Flask. The client sends the message history and a model identifier to a dedicated endpoint (e.g., \texttt{/api/chat}), which then forwards the request to the model provider and returns only the generated text to the browser. Critically, API credentials are stored exclusively on the server and are never embedded in client-side code, preventing potential exposure of keys in the participant’s browser environment. Centralizing model access through the proxy also supports reproducibility by enforcing consistent defaults (e.g., model selection when unspecified) and enables future extensions such as request auditing, rate limiting, and standardized retry handling.

\subsection{Data Storage Layer}
To support the analysis of expert interaction traces, the system logs study and simulation data to a cloud database (Firestore). Logging occurs at both the session level (participant background responses and scenario configuration) and the turn level (generated PLWD responses, expert ratings/comments, generated caregiver strategy options, and the final caregiver response submitted). This structure enables downstream aggregation by dementia stage, care setting, ADL, and turn index, and facilitates the export of complete transcripts (e.g., TXT/CSV) for archival and offline analysis.  During deployment, Firestore access is restricted to the minimal set of operations required for logging and export.   Where applicable, database rules are configured to enforce authenticated access and prevent unintended reads of other participants’ records.

\subsection{Privacy and Security Considerations}
The system is designed to minimize the exposure of sensitive information. First, model API keys are protected via a backend proxy and are never transmitted to or stored in the browser. Second, the participant questionnaire is limited to study-relevant, dementia-caregiving background variables (e.g., professional role and dementia-care experience) and it does not require direct identifiers such as names, phone numbers, or addresses. Third, participant sessions are associated with pseudonymous identifiers (e.g., automatically generated \texttt{Guest\_XXXXX} IDs) to support longitudinal trace grouping without storing any personal identity information. Finally, exported transcripts are generated on-demand for participants and contain only the dialogue and study metadata captured during the session, enabling transparency while maintaining a minimal-data collection posture.

\section{User Study for Simulator Evaluation with Domain Experts}
We conducted a formative, \emph{domain-expert-in-the-loop} evaluation to enhance the simulator workflow and refine core components prior to broader data collection and future extensions toward autonomous patient--caregiver co-simulation. This phase was designed to (1) identify usability issues in the web interface (e.g., clarity of instructions, turn progression, rating flow, and caregiver response editing), and (2) surface systematic failure modes in LLM-generated PLWD behavior (e.g., dementia-stage incongruence, unrealistic ADL progression, or implausible emotional responses). We treat this study as formative rather than confirmatory; its purpose is to assess feasibility and perceived realism via turn-based expert ratings and qualitative feedback, rather than to support clinical deployment or population-level generalization claims.

Participants were recruited via email and online social media and included dementia-care stakeholders with diverse backgrounds (e.g., clinicians, nurses, educators, researchers, and students with prior dementia-care knowledge). No compensation was provided. Inclusion criteria were limited to age $\geq$ 18 years old and willing to participate, as the study aimed to elicit formative usability and realism feedback rather than hypothesis testing. To preserve privacy, we did not collect direct personally identifiable information (PII) such as names, email addresses, phone numbers, or institutional identifiers. Instead, participants were assigned randomly generated pseudonymous identifiers used solely to associate responses within a session. 

We collected only non-identifying background variables related to dementia-care experience (e.g., gender, age range, education level, country/region of work, and other dementia-care background) to contextualize feedback. All logged interaction traces---including scenario selections, generated PLWD responses, turn-level realism ratings (1--5) with optional comments, and submitted caregiver responses---were stored and analyzed using these pseudonymous identifiers. The study protocol was approved by the authors' Institutional Review Board (IRB \#\#\#\#); to preserve double-blind review, we omit the protocol number and institutional identifiers in this submission.


\begin{figure*}[ht!]
  \centering  \includegraphics[width=1.3\columnwidth]{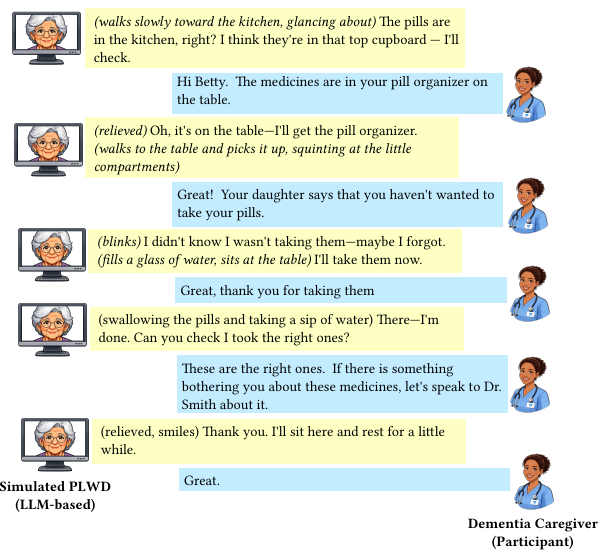}
\caption{An example of PLWD-caregiver interaction during taking medicine in PLWD's home. The PLWD is with middle-stage Alzheimer's dementia, living at their home for more than 1 year. The text in parentheses represents PLWD's nonverbal behavior. Other text represent PLWD's and caregiver's verbal response, correspondingly
}
 \label{fig:caregiver}
\end{figure*}

\section{Results and Discussion}
\subsection{Demographics and Study Coverage}
We collected interaction traces from 14 unique participants across 18 simulation sessions, yielding 112 turn-level PLWD responses with associated realism ratings. Most participants completed a single session, while 3 participants ran multiple sessions with different scenario parameters, illustrating that the simulator supported repeated testing across conditions. Participants reported a range of backgrounds relevant to dementia care and communication, as summarized in Table \ref{tab:demographics}; among respondents who provided demographics, the most common reported age range was 35–44 and the most commonly reported gender was female (n = 11). Education level skewed toward advanced training, with many respondents reporting doctoral-level education. Domain-relevant expertise was common: based on the presence of follow-up responses in the branching survey (i.e., the items shown only after a “Yes” response), multiple participants indicated prior experience caring for someone with dementia and/or working as healthcare professionals, and many reported moderate-to-high familiarity with dementia communication strategies. We report demographic distributions as counts per item (Table \ref{tab:demographics}) and explicitly note missingness where participants did not provide optional fields; all participants were logged via pseudonymous IDs, and no direct personally identifying information was collected.


\begin{table}[t]
\caption{Participant characteristics for the online evaluation with dementia care experts of the PLWD simulator ($N=14$). Values are reported as n (\%).}
\label{tab:demographics}
\centering
\renewcommand{\arraystretch}{1.15}
\setlength{\tabcolsep}{4pt} 

\begin{tabularx}{\columnwidth}{@{}
  >{\raggedright\arraybackslash}p{0.35\columnwidth}
  >{\raggedright\arraybackslash}X
  >{\raggedleft\arraybackslash}p{1.85cm}
@{}}
\toprule
\textbf{Characteristic} & \textbf{Category} & \textbf{n (\%)}\\
\midrule

\multirow[t]{6}{0.25\columnwidth}{Age range}
  & 18--24  & 1 (7.1\%)\\
  & 25--34  & 1 (7.1\%)\\
  & 35--44  & 6 (42.9\%)\\
  & 45--54  & 5 (35.7\%)\\
  & 55--64  & 1 (7.1\%)\\
  & 65+     & 0 (0.0\%)\\
\midrule

\multirow[t]{2}{0.35\columnwidth}{Gender}
  & Male    & 3 (21.4\%)\\
  & Female  & 11 (78.6\%)\\
\midrule

\multirow[t]{2}{0.35\columnwidth}{Education}
  & Master's  & 2 (14.3\%)\\
  & Doctorate & 12 (85.7\%)\\
\midrule

\multirow[t]{5}{0.35\columnwidth}{Occupation (if applicable, select all)}
  & Researcher & 10 (71.4\%)\\
  & Clinician  & 7 (50.0\%)\\
  & Nurse      & 4 (28.6\%)\\
  & Teacher    & 7 (50.0\%)\\
  & Student    & 1 (7.1\%)\\
\midrule

\multirow[t]{4}{0.35\columnwidth}{Dementia care role (if applicable, select all)}
  & Family caregiver       & 1 (7.1\%)\\
  & Professional caregiver & 2 (14.3\%)\\
  & Healthcare provider    & 6 (42.9\%)\\
  & None                   & 7 (50.0\%)\\
\midrule

\multirow[t]{5}{0.35\columnwidth}{Formal training}
  & Basic training/workshop & 2 (14.3\%)\\
  & Certificate             & 3 (21.4\%)\\
  & Degree program          & 3 (21.4\%)\\
  & Continuing education    & 5 (35.7\%)\\
  & None                    & 1 (7.1\%)\\
\midrule

\multirow[t]{4}{0.35\columnwidth}{Dementia communication strategy familiarity}
  & Slightly     & 2 (14.3\%)\\
  & Moderately   & 1 (7.1\%)\\
  & Very         & 10 (71.4\%)\\
  & Expert level & 1 (7.1\%)\\

\bottomrule
\end{tabularx}
\end{table}

\begin{figure}[ht!]
  \centering  \includegraphics[width=0.8\columnwidth]{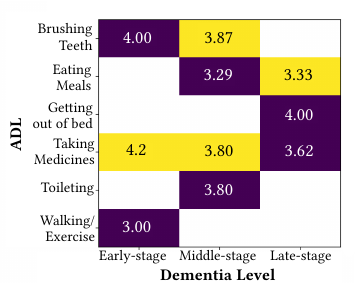}
\caption{Mean participant-rated realism (1–5 scale) of generated PLWD responses across ADL and dementia stage. Text in each cell shows the average turn-level realism rating provided by participants for the specified ADL × dementia stage combination. Cell shading indicates the number of rated occurrences contributing to the mean (yellow: 2 occurrences; purple: 1 occurrence). Blank cells indicate combinations not evaluated during the study.}
 \label{fig:adlxdementia}
\end{figure}

\subsection{Dementia-related ADL Simulation Parameter Selection}
Figure~\ref{fig:adlxdementia} illustrates how evaluators sampled the scenario space across ADL × dementia stage, where cell shading encodes how often a combination was selected (one vs. two sessions) and the number printed in each cell is the mean realism rating for the PLWD’s responses under that configuration. Overall, selections were sparse and concentrated, with most ADL–stage pairs evaluated only once and several not evaluated at all, which is typical of formative studies where participants self-select scenarios. The most frequently tested ADL was Taking Medicines, which was sampled in early-stage (mean 4.2; 2 occurrences), middle-stage (mean 3.80; 2 occurrences), and late-stage (mean 3.62; 1 occurrence), making it the only ADL spanning all three stages and suggesting that medication assistance is a salient, familiar, and practically meaningful context for expert evaluation. Brushing Teeth was also selected across multiple sessions, with early-stage rated highly (mean 4.00; 1 occurrence) and middle-stage remaining comparably strong (mean 3.87; 2 occurrences), indicating that the simulator’s behavior for structured hygiene routines may be perceived as relatively stage-congruent and believable. In contrast, Eating Meals received lower mid-range ratings in both middle-stage (mean 3.29; 1 occurrence) and late-stage (mean 3.33; 2 occurrences), which may reflect that meal contexts demand more nuanced modeling of appetite, attention, sequencing, and assistance needs—i.e., multiple interacting factors that can expose weaknesses in task grounding and stage-appropriate impairment cues. Late-stage coverage was otherwise limited but notably included Getting out of bed (mean 4.00; 1 occurrence), suggesting that, at least in this sampled instance, the simulator produced a plausibly impaired but coherent response pattern for a high-assistance mobility scenario. Remaining combinations were sampled only once, including Toileting × middle-stage (mean 3.80) and Walking/Exercise × early-stage (mean 3.00); the comparatively lower rating for walking/exercise may indicate a mismatch between the model’s response style and expectations for early-stage autonomy and conversational naturalness in less-structured activities. Taken together, the figure highlights two complementary findings: (i) where experts chose to test (a practical “stress test” focus on medication and mid-stage scenarios), and (ii) where perceived realism appears strongest vs. weaker in this initial dataset—while also underscoring the need for broader, more balanced sampling across ADLs and stages in subsequent rounds to draw stronger comparative conclusions.

\begin{figure}[ht!]
  \centering  \includegraphics[width=0.9\columnwidth]{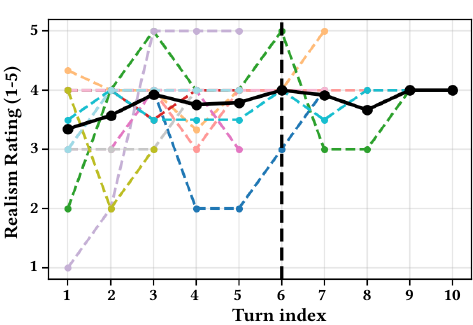}
\caption{Turn-by-turn realism ratings for the LLM-generated PLWD responses. Each colored dashed line corresponds to one simulation session (18 sessions total), showing the participant’s realism rating (1–5) at each turn. The thick black line plots the mean realism rating at each turn across sessions. The vertical dashed black line marks the typical stopping point (median session length).}
 \label{fig:turnvsrate}
\end{figure}

\subsection{Participant Ratings and Responses}
Figure~\ref{fig:turnvsrate} summarizes participants’ turn-level realism ratings of the LLM-generated PLWD responses, collected on a 5-point Likert scale based on whether each response matched the selected dementia stage in the chosen scenario. Overall, ratings indicate moderate-to-high perceived plausibility: the across-participant mean (thick black trace) remains in the upper mid-range ($\approx$ 3.5–4.0) across the interaction, with a small increase over the first few turns (Turns 1-3) and no clear monotonic degradation thereafter among sessions that continue. At the same time, the colored participant traces reveal substantial heterogeneity, including occasional sharp dips (e.g., ratings near 2 in mid-turns), which suggests localized breakdowns rather than uniform decline—consistent with scenario- or context-dependent failures such as stage-incongruent language/initiative or task-grounding errors (e.g., implausible ADL step progression). The vertical dashed line marks the typical stopping point (median session length $\approx$ 6 turns), indicating that many interactions concluded within $\approx$ 6 exchanges; importantly, ratings in later turns reflect fewer continuing sessions and should be interpreted with this survivorship effect in mind (i.e., later-turn averages may be biased toward interactions that remained coherent enough for participants to continue).

\begin{table*}[t]
\centering
\small
\setlength{\tabcolsep}{4pt}
\renewcommand{\arraystretch}{1.15}
\caption{Failure-mode taxonomy from optional free-text clinician comments. Comments were coded at the turn level and a turn may receive multiple codes; therefore, frequencies sum to more than $N=20$ commented turns. Frequency is reported as \# commented turns (percent of commented turns). Example excerpts are lightly edited for brevity.}
\label{tab:failure_modes}
\begin{tabularx}{\textwidth}{X X X c X}
\toprule
\textbf{Failure mode} & \textbf{Definition} & \textbf{Example excerpt} & \textbf{Freq.} & \textbf{Design implication / mitigation} \\
\midrule
Stage mismatch &
Comment suggests the response does not match the selected dementia stage (too coherent/able or too impaired/disoriented). &
Expected more resistance; response seems closer to early-stage dementia. &
5 (25\%) & Tighten stage constraints (memory/orientation/language) and cap planning/initiative; add guardrails for severe stages. \\
Task/ADL grounding error &
Comment flags implausible ADL steps, sequencing, assumptions about task setup, or multi-step planning beyond expected ability. &
Multi-step ADL planning (e.g., wiping + tucking napkin) may exceed moderate-stage capacity. &
9 (45\%) & Refine ADL step plans and enforce step-local responses; add object-location context and prevent skipping without caregiver guidance. \\
Care-setting mismatch &
Comment indicates the PLWD behavior depends on or contradicts the selected environment (e.g., home vs facility; where objects are stored). &
Assisted-living workflows: medications are centralized (not stored with the resident). &
4 (20\%) & Add setting-specific constraints (storage policies, routines) and use duration to modulate familiarity; avoid inconsistent assumptions. \\
Overcompliance / low resistance &
Comment suggests the simulated PLWD is too agreeable/compliant and should show more refusal, anxiety, or resistance. &
PLWD is too agreeable given selected challenges (distraction/anxiety/refusal). &
5 (25\%) & Increase probability of hesitation/refusal in moderate/late stages and when challenges are present; require repeated cues before progress. \\
Language/utterance unnaturalness &
Comment critiques phrasing (overly well-formed, too long/complex, inappropriate word choice) relative to expected communication ability. &
Utterance is overly well-structured/verbose; clinicians suggested shorter, simpler phrasing. &
5 (25\%) & Constrain to simpler phrasing; discourage formal grammar; avoid infantilizing terms; permit more nonverbal actions instead of naming objects. \\
Needs more prompting / support &
Comment suggests the scenario should require repeated prompting, nudging, or more caregiver assistance across turns. &
Even when plausible, progress often requires nudging and repeated reminding across turns. &
4 (20\%) & Prompt the PLWD model to stall/ask for clarification; progress only with repeated caregiver prompts in moderate/late stages. \\
\bottomrule
\end{tabularx}
\end{table*}

\begin{figure}[ht!]
  \centering  \includegraphics[width=\columnwidth]{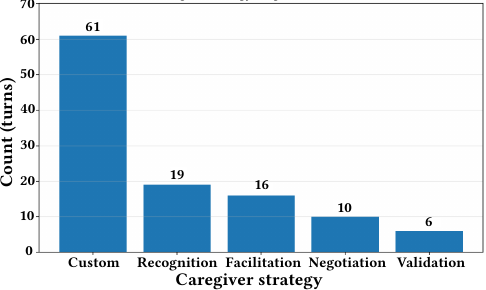}
\caption{Caregiver response strategy usage across all turns. “Custom” denotes participant-authored free-text replies or any strategy suggestion that was selected and subsequently edited before submission.}
 \label{fig:caregiver}
\end{figure}
\subsection{Caregiver Strategy Usage}
Figure~\ref{fig:caregiver} reports how participants responded as caregivers, either by selecting one of four strategy-scaffolded options or composing a custom response. Custom responses were most common (61/112 turns, 54.5\%), indicating that participants frequently preferred to author their own caregiver utterances rather than adopt an LLM-generated template—an observation aligned with clinician feedback that suggested options could be overly verbose for practical dementia-care interactions. Among scaffolded strategies, Recognition (19/112, 17.0\%) and Facilitation (16/112, 14.3\%) were used most often, reflecting the task-oriented nature of ADL assistance where rapport-building and stepwise support are central (e.g., brief, encouraging prompts that reduce cognitive load, "I'll open it and show you today's pills. Let's sit at the table and get a glass of water first"). Negotiation (10/112, 8.9\%) and Validation (6/112, 5.4\%) were less frequently selected; qualitatively, negotiation options sometimes introduced additional wording or choice structures that participants perceived as unnecessarily long in-the-moment (eg. ``Let's check the label together. Would you like me to read the name or open it and show today's pills?"), while validation may have been less applicable in scenarios that were primarily procedural rather than emotionally escalated. Together, these patterns suggest that the strategy scaffold is useful as a structure, but that brevity and editability are key design requirements—especially if suggested responses are intended to be directly usable in fast-paced caregiving exchanges.

\subsection{Qualitative Analysis}
We performed a lightweight thematic analysis of the optional free-text comments provided alongside turn-level realism ratings and iteratively developed a codebook of $N=6$ failure modes (Table~\ref{tab:failure_modes}). Comments were coded at the turn level; because a single comment could reference multiple issues (e.g., stage incongruence and over-compliance), a turn could receive multiple codes. We report the frequency of each code across commented turns and the mean realism rating for turns where each code was present.

Clinician comments provided targeted engineering insight into where the simulator breaks and what to prioritize for refinement (Table~\ref{tab:failure_modes}). Comments were relatively sparse (20 commented turns), but they concentrated on lower-quality outputs: turns with comments received lower realism ratings on average than turns without comments, suggesting participants primarily used free-text feedback to document breakdowns rather than to restate satisfactory performance. The most frequent theme was \textit{task/ADL grounding error} (9/20 commented turns, 45\%), and it was associated with lower ratings (mean 2.67), indicating that the model often struggled to maintain plausible stepwise progression (e.g., planning several ADL actions ahead or assuming task setup that had not been established by the caregiver). This points to prompt and scaffolding changes that enforce \emph{step-local} behavior (responding primarily about the current/next step), and to strengthening the ADL progress representation used to ground generation. Although less frequent, \textit{care-setting mismatch} (4/20, mean 2.00) produced the lowest ratings, reflecting high-severity errors where responses contradicted setting-specific norms (e.g., medication storage and routines in assisted living). This motivates adding explicit setting constraints (and using care-duration to modulate familiarity) to prevent incorrect environmental assumptions. \textit{Overcompliance} (5/20, mean 2.80) and \textit{stage mismatch} (5/20, mean 3.00) suggest the simulator can be overly cooperative and cognitively capable even when refusal/anxiety or more severe impairment is selected, motivating stronger stage-conditioned guardrails (caps on initiative/planning, increased hesitation/refusal probability, and greater reliance on caregiver prompting). Finally, \textit{language unnaturalness} (5/20, mean 3.00) highlights an interactional realism issue: clinicians preferred shorter, simpler phrasing and fewer explicit object labels, consistent with dementia-appropriate communication. Notably, \textit{needs more prompting} (4/20, mean 3.75) tended to appear even when responses were otherwise plausible, indicating that improving realism is not only about correcting errors but also about modeling realistic delays and repeated cueing during ADLs. Together, these themes provide a concrete refinement loop: prioritize high-impact setting constraints, improve task grounding and step-wise progression, and tighten stage- and style-specific generation constraints.

\subsection{Limitations and Future Work}
This study is formative and has several limitations that bound interpretation. First, coverage of the scenario space was sparse and participant-driven: most ADL $\times$ stage configurations were evaluated in only one session (Fig.~\ref{fig:adlxdementia}), and several combinations were not sampled. This self-selection is appropriate for early stress-testing, but it limits comparative claims across ADLs, stages, and care settings. Second, the sample size was modest ($N=14$ participants; 18 sessions) and optional qualitative feedback was relatively sparse (20 commented turns). While comments concentrated on lower-rated turns and yielded actionable failure modes (Table~\ref{tab:failure_modes}), a larger dataset with broader commenting would support more stable estimates of failure-mode prevalence and stronger conclusions about which errors dominate under which conditions. Third, turn-level trends should be interpreted cautiously due to session-length variability and survivorship effects: later turns reflect fewer continuing sessions (Fig.~\ref{fig:turnvsrate}), which may bias later-turn averages toward interactions that remained coherent enough for participants to continue.

These limitations motivate several next steps. On the evaluation side, we will pursue broader and more balanced sampling across ADLs, dementia stages, and care contexts, including protocols that encourage systematic coverage and richer qualitative annotation. On the system side, we will strengthen ADL task grounding by enforcing step-local generation tied to an explicit representation of current/next ADL progress, and by adding stricter care-setting constraints (e.g., assisted living vs.\ home routines) to reduce environment-inconsistent assumptions. We will also tighten stage-conditioned guardrails to better calibrate cognitive capability and initiative (e.g., limiting multi-step planning, increasing realistic hesitation/refusal, and modeling repeated prompting), and improve style control to yield shorter, dementia-appropriate utterances. Finally, we will explore using the collected ratings and critiques as supervision signals for iterative prompt refinement and, where appropriate, model adaptation, and we will extend the platform toward validated patient--caregiver co-simulation to support scalable data generation, benchmarking, and policy learning for assistive AI and robotics in dementia care.

\section{Conclusion}
In this work, we introduced an interactive, web-based simulator that uses an LLM to generate multi-turn behaviors for a person living with Alzheimer’s dementia during activities of daily living (ADLs). The simulator supports expert-configured scenarios (dementia stage and care context), elicits turn-level realism ratings with optional critique, and enables caregiver responses via either free-text authoring or editable strategy-scaffolded suggestions. 
In a formative, domain-expert-in-the-loop evaluation (14 participants; 18 sessions; 112 rated turns), experts with dementia care experience judged the generated PLWD behaviors as moderately to highly plausible overall, while also providing concrete, turn-level feedback that reveals where and how realism breaks in practice.
By pairing controllable simulation parameters with structured expert evaluation, this work offers a practical pathway for assessing stage- and context-congruent LLM-based patient simulation and for iteratively improving simulation fidelity in safety- and privacy-sensitive dementia-care setting. Ultimately, our work lays groundwork for more reliable simulated interactions that can support the development and evaluation of assistive technologies and socially assistive AI \& robots for dementia care.

\bibliographystyle{ACM-Reference-Format}
\bibliography{references}

@article{AA2024diseasefact,
  title={2024 Alzheimer's Disease Facts and Figures},
  author={{Alzheimer's Association}},
  journal={Alzheimer's \& Association},
  volume={20},
  number={5},
  pages={3708--3821},
  year={2024},
  month={May},
  doi={10.1002/alz.13809},
  eprint={PMC11095490},
  pmid={38689398}
}

@article{langbaum2023recommendations,
  title={Recommendations to address key recruitment challenges of Alzheimer's disease clinical trials},
  author={Langbaum, Jessica B and Zissimopoulos, Julie and Au, Rhoda and Bose, Niranjan and Edgar, Chris J and Ehrenberg, Evan and Fillit, Howard and Hill, Carl V and Hughes, Lynne and Irizarry, Michael and others},
  journal={Alzheimer's \& Dementia},
  volume={19},
  number={2},
  pages={696--707},
  year={2023},
  publisher={Wiley Online Library}
}

@article{giebel2014deterioration,
  title={Deterioration of basic activities of daily living and their impact on quality of life across different cognitive stages of dementia: a European study},
  author={Giebel, Clarissa M and Sutcliffe, Caroline and Stolt, Minna and Karlsson, Staffan and Renom-Guiteras, Anna and Soto, Maria and Verbeek, Hilde and Zabalegui, Adelaida and Challis, David},
  journal={International psychogeriatrics},
  volume={26},
  number={8},
  pages={1283--1293},
  year={2014},
  publisher={Cambridge University Press}
}

@article{yin2023risk,
  title={Risk factors associated with home care safety for older people with dementia: family caregivers’ perspectives},
  author={Yin, Guo and Lin, Siting and Chen, Linghui},
  journal={BMC geriatrics},
  volume={23},
  number={1},
  pages={224},
  year={2023},
  publisher={Springer}
}

@article{spira2006behavioral,
  title={Behavioral interventions for agitation in older adults with dementia: an evaluative review},
  author={Spira, Adam P and Edelstein, Barry A},
  journal={International psychogeriatrics},
  volume={18},
  number={2},
  pages={195--225},
  year={2006},
  publisher={Cambridge University Press}
}

@article{ghafurian2021social,
  title={Social robots for the care of persons with dementia: a systematic review},
  author={Ghafurian, Moojan and Hoey, Jesse and Dautenhahn, Kerstin},
  journal={ACM Transactions on Human-Robot Interaction (THRI)},
  volume={10},
  number={4},
  pages={1--31},
  year={2021},
  publisher={ACM New York, NY, USA}
}

@misc{holderautomated,
  title={Automated activity-aware prompting for activity initiation. Gerontechnology 11 (4), 534--544 (2013)},
  author={Holder, LB and Cook, DJ}
}

@article{pappada2021assistive,
  title={Assistive technologies in dementia care: an updated analysis of the literature},
  author={Pappad{\`a}, Alessandro and Chattat, Rabih and Chirico, Ilaria and Valente, Marco and Ottoboni, Giovanni},
  journal={Frontiers in psychology},
  volume={12},
  pages={644587},
  year={2021},
  publisher={Frontiers Media SA}
}

@inproceedings{park2023generative,
  title={Generative agents: Interactive simulacra of human behavior},
  author={Park, Joon Sung and O'Brien, Joseph and Cai, Carrie Jun and Morris, Meredith Ringel and Liang, Percy and Bernstein, Michael S},
  booktitle={Proceedings of the 36th annual acm symposium on user interface software and technology},
  pages={1--22},
  year={2023}
}

@article{argyle2023out,
  title={Out of one, many: Using language models to simulate human samples},
  author={Argyle, Lisa P and Busby, Ethan C and Fulda, Nancy and Gubler, Joshua R and Rytting, Christopher and Wingate, David},
  journal={Political Analysis},
  volume={31},
  number={3},
  pages={337--351},
  year={2023},
  publisher={Cambridge University Press}
}

@article{hu2025simbench,
  title={Simbench: Benchmarking the ability of large language models to simulate human behaviors},
  author={Hu, Tiancheng and Baumann, Joachim and Lupo, Lorenzo and Collier, Nigel and Hovy, Dirk and R{\"o}ttger, Paul},
  journal={arXiv preprint arXiv:2510.17516},
  year={2025}
}

@article{huang2025survey,
  title={A survey on hallucination in large language models: Principles, taxonomy, challenges, and open questions},
  author={Huang, Lei and Yu, Weijiang and Ma, Weitao and Zhong, Weihong and Feng, Zhangyin and Wang, Haotian and Chen, Qianglong and Peng, Weihua and Feng, Xiaocheng and Qin, Bing and others},
  journal={ACM Transactions on Information Systems},
  volume={43},
  number={2},
  pages={1--55},
  year={2025},
  publisher={ACM New York, NY}
}

@article{qu2024performance,
  title={Performance and biases of large language models in public opinion simulation},
  author={Qu, Yao and Wang, Jue},
  journal={Humanities and Social Sciences Communications},
  volume={11},
  number={1},
  pages={1--13},
  year={2024},
  publisher={Palgrave}
}

@article{brugge2024large,
  title={Large language models improve clinical decision making of medical students through patient simulation and structured feedback: a randomized controlled trial},
  author={Br{\"u}gge, Emilia and Ricchizzi, Sarah and Arenbeck, Malin and Keller, Marius Niklas and Schur, Lina and Stummer, Walter and Holling, Markus and Lu, Max Hao and Darici, Dogus},
  journal={BMC medical education},
  volume={24},
  number={1},
  pages={1391},
  year={2024},
  publisher={Springer}
}

@inproceedings{sehgal2025pal,
  title={PAL: Designing Conversational Agents as Scalable, Cooperative Patient Simulators for Palliative-Care Training},
  author={Sehgal, Neil KR and Kambhamettu, Hita and Chang, Allen and Zhu, Andrew and Ungar, Lyle and Guntuku, Sharath Chandra},
  booktitle={Companion Publication of the 2025 Conference on Computer-Supported Cooperative Work and Social Computing},
  pages={346--350},
  year={2025}
}

@article{tanana2019development,
  title={Development and evaluation of ClientBot: Patient-like conversational agent to train basic counseling skills},
  author={Tanana, Michael J and Soma, Christina S and Srikumar, Vivek and Atkins, David C and Imel, Zac E},
  journal={Journal of medical Internet research},
  volume={21},
  number={7},
  pages={e12529},
  year={2019},
  publisher={JMIR Publications Toronto, Canada}
}

@article{pilnick2023conversation,
  title={Conversation Analysis Based Simulation (CABS): A method for improving communication skills training for healthcare practitioners},
  author={Pilnick, Alison and O'Brien, Rebecca and Beeke, Suzanne and Goldberg, Sarah and Murray, Megan and Harwood, Rowan H},
  journal={Health Expectations},
  volume={26},
  number={6},
  pages={2461--2474},
  year={2023},
  publisher={Wiley Online Library}
}

@article{cook2025virtual,
  title={Virtual patients using large language models: Scalable, contextualized simulation of clinician-patient dialogue with feedback},
  author={Cook, David A and Overgaard, Joshua and Pankratz, V Shane and Del Fiol, Guilherme and Aakre, Chris A},
  journal={Journal of Medical Internet Research},
  volume={27},
  pages={e68486},
  year={2025},
  publisher={JMIR Publications Toronto, Canada}
}

@article{luo2025large,
  title={A large language model digital patient system enhances ophthalmology history taking skills},
  author={Luo, Ming-Jie and Bi, Shaowei and Pang, Jianyu and Liu, Lixue and Tsui, Ching-Kit and Lai, Yunxi and Chen, Wenben and Yang, Yahan and Xu, Kezheng and Zhao, Lanqin and others},
  journal={NPJ Digital Medicine},
  volume={8},
  number={1},
  pages={502},
  year={2025},
  publisher={Nature Publishing Group UK London}
}

@inproceedings{zhu2025designing,
  title={Designing VR Simulation System for Clinical Communication Training with LLMs-Based Embodied Conversational Agents},
  author={Zhu, Xiuqi Tommy and Cheerman, Heidi and Cheng, Minxin and Kiami, Sheri R and Chukoskie, Leanne and McGivney, Eileen},
  booktitle={Proceedings of the Extended Abstracts of the CHI Conference on Human Factors in Computing Systems},
  pages={1--9},
  year={2025}
}

@report{alz_facts_figures_2025,
  author       = {{Alzheimer's Association}},
  title        = {2025 Alzheimer's Disease Facts and Figures},
  year         = {2025},
  institution  = {Alzheimer's Association},
  url          = {https://www.alz.org/getmedia/ef8f48f9-ad36-48ea-87f9-b74034635c1e/alzheimers-facts-and-figures.pdf},
  note         = {Accessed: 2026-02-12}
}

@online{mayo_alzheimers_stages_2025,
  author       = {{Mayo Clinic Staff}},
  title        = {Alzheimer's stages: How the disease progresses},
  year         = {2025},
  month        = may,
  day          = {9},
  organization = {Mayo Clinic},
  url          = {https://www.mayoclinic.org/diseases-conditions/alzheimers-disease/in-depth/alzheimers-stages/art-20048448},
  note         = {Accessed: 2026-02-12}
}

@article{savundranayagam2015language,
  title={Language-based communication strategies that support person-centered communication with persons with dementia},
  author={Savundranayagam, Marie Y and Moore-Nielsen, Kelsey},
  journal={International Psychogeriatrics},
  volume={27},
  number={10},
  pages={1707--1718},
  year={2015},
  publisher={Cambridge University Press}
}

@article{o2022expanding,
  title={Expanding the conversation: A person-centred communication enhancement model},
  author={O’Rourke, Deanne J and Lobchuk, Michelle M and Thompson, Genevieve N and Lengyel, Christina},
  journal={Dementia},
  volume={21},
  number={5},
  pages={1596--1617},
  year={2022},
  publisher={SAGE Publications Sage UK: London, England}
}

\end{document}